\documentclass[a4paper]{jpconf}
\usepackage{graphicx}
\usepackage{multirow}
\usepackage{amsmath,amssymb,upgreek}
\usepackage[bottom]{footmisc}
\usepackage{tikz}
\usetikzlibrary{external}
\usetikzlibrary{decorations.pathmorphing}
\usetikzlibrary{calc}
\usetikzlibrary{positioning}
\usepackage{feynmp}
\usepackage{siunitx}

\usepackage{xcolor}
\usepackage{lipsum}

\usepackage{todonotes}

\begin{document}
\title{Searching for New Physics with DarkLight at the ARIEL Electron-Linac}

\author{The~DarkLight~Collaboration, E~Cline$^1$, R~Corliss$^1$, J~C~Bernauer$^{1,2}$, R~Alarcon$^3$, R~Baartman$^4$, S~Benson$^5$, J~Bessuille$^6$, D~Ciarniello$^4$, A~Christopher$^7$, A~Colon$^1$, W~Deconinck$^8$, K~Dehmelt$^1$, A~Deshpande$^1$, J~Dilling$^4$, D~H~Dongwi$^9$, P~Fisher$^6$, T~Gautam$^7$, M~Gericke$^8$, D~Hasell$^6$, M~Hasinoff$^{10}$, E~Ihloff$^6$, R~Johnston$^6$, R~Kanungo$^{11}$, J~Kelsey$^6$, O~Kester$^4$, M~Kohl$^7$, I~Korover$^6$, R~Laxdal$^4$, S~Lee$^6$, X~Li$^6$, C~Ma$^1$, A~Mahon$^4$, J~W~Martin$^{12}$, R~Milner$^6$, M~Moore$^6$, P~Moran$^6$, J~Nazeer$^7$, K~Pachal$^4$, T~Patel$^7$, T~Planche$^4$, M~Rathnayake$^7$, M~Suresh$^7$, C~Vidal$^6$, Y~Wang$^6$, S~Yen$^4$}

\address{$^1$ Department of Physics and Astronomy, Stony Brook University, Nicolls Road, Stony Brook, NY, 11794, USA}
\address{$^2$ RIKEN BNL Research Center, Brookhaven National Laboratory, 20 Brookhaven Ave, Upton, NY, 11973, USA}
\address{$^3$ Department of Physics, Arizona State University, 550 E Tyler Drive, Tempe, AZ, 85287, USA}
\address{$^4$ TRIUMF, 4004 Wesbrook Mall, Vancouver, BC, V6T 2A3, CA}
\address{$^5$ Thomas Jefferson National Accelerator Facility, 12000 Jefferson Avenue, Newport News, Virginia, 23606, USA}
\address{$^6$ Laboratory for Nuclear Science, Massachusetts Institute of Technology, 77 Massachusetts Avenue, Cambridge, MA 02139, USA}
\address{$^7$ Department of Physics, Hampton University, 268 Emancipation Drive, Hampton, VA, 23668, USA}
\address{$^8$ Department of Physics and Astronomy, University of Manitoba, 30a Sifton Road, Winnipeg, MB, R3T 2N2, CA}
\address{$^9$ Lawrence Livermore National Laboratory, 7000 East Ave, Livermore, CA 94550, USA}
\address{$^{10}$ Department of Physics and Astronomy, University of British Columbia, 6224 Agricultural Road, Vancouver, BC V6T 1Z1, CA}
\address{$^{11}$ Department of Physics and Astronomy, Saint Mary's University, 923 Robie Street, Halifax, NS, B3H 3C3, CA}
\address{$^{12}$ Department of Physics, University of Winnipeg, 515 Portage Avenue, Winnipeg, MB, R3B 2E9, CA}

\ead{ethan.cline@stonybrook.edu}

\begin{abstract}
The search for a dark photon holds considerable interest in the physics community. Such a force carrier would begin to illuminate the dark sector. Many experiments have searched for such a particle, but so far it has proven elusive. In recent years the concept of a low mass dark photon has gained popularity in the physics community. Of particular recent interest is the $^8$Be and $^4$He anomaly, which could be explained by a new fifth force carrier with a mass of 17 MeV/$c^2$. The proposed Darklight experiment would search for this potential low mass force carrier at ARIEL in the 10-20 MeV/$c^2$ e$^+$e$^-$ invariant mass range. This proceeding will focus on the experimental design and physics case of the Darklight experiment.
\end{abstract}

\section{Introduction}

Recently, there has been a focus in both experimental and theoretical physics communities on a mediator of a new fifth force with mass lower than 1~GeV/$c^2$. 
In addition to cosmological motivations, observed anomalies in measurements involving the muon and nuclear transitions hint at this possibility.
For example, the observed 4.2$\sigma$ deviation between the measured and expected anomalous magnetic moment of the muon~\cite{Abi2021}
can be explained by a fifth force mediator with mass in the range 10 to 100~MeV~\cite{Frey2009}. 
The search for a dark photon has been extensively covered through the study of $\pi^0$-decay and much of the parameter space of coupling and mass that corresponds to these anomalies is excluded at 2$\sigma$~\cite{NA642020}. 
However, it is not required that a potential dark sector coupling must be directly proportional to the electric charges, so a more general fifth force can not yet be ruled out.

It is possible to adjust the quark couplings of a fifth force to satisfy existing constraints and still allow such a force acting via lepton coupling to be experimentally detected. In addition to the muon g-2, other recently-reported experimental signatures motivate focused searches for a fifth force carrier at low energies. A group studying the decays of excited states of $^8$Be and $^4$He to their ground state have found a 6.8$\sigma$ anomaly in the opening angle and invariant mass distribution of $e^+e^-$ pairs produced in these transitions~\cite{Kra2016,Kra2019a,Kra2019} (The ATOMKI anomalies).  While these discrepancies may be the result of experimental effects or unidentified nuclear transitions they are also consistent with the production of a new boson with a mass around 17~MeV/$c^2$ (the X17 particle). 

New bosons that couple atomic electrons with neutrons in the nucleus are also implicated in atomic physics experiments. The effect of this new interaction on energy levels and transition frequencies could be detected through precision isotope shift measurements.  In particular, the scaled isotope shifts of two different transitions should exhibit a linear relationship (the so-called {\it King plot}).  A deviation from linearity can be evidence of a new force mediator.  Such deviations at the 3$\sigma$ level have been reported~\cite{Counts2020} in the isotope shifts for five Yb$^+$ isotopes on two narrow optical quadrupole transitions $^2$S$_{1/2} \rightarrow {}^2$D$_{3/2} \rightarrow {}^2$D$_{5/2}$.

Motivated by these developments, we have designed the DarkLight experiment to use the future \SI{50}{\mega\electronvolt} electron beam from the e-linac driver at the Advanced Rare IsotopE Laboratory (ARIEL) to search for evidence of the reported ATOMKI anomaly in $e^+e^-$ final-states.

\subsection{Beyond the Standard Model Interpretations}
\begin{figure}[tb]
{\centering
\includegraphics[width=0.95\textwidth,clip,trim=0 3cm 0 3cm]{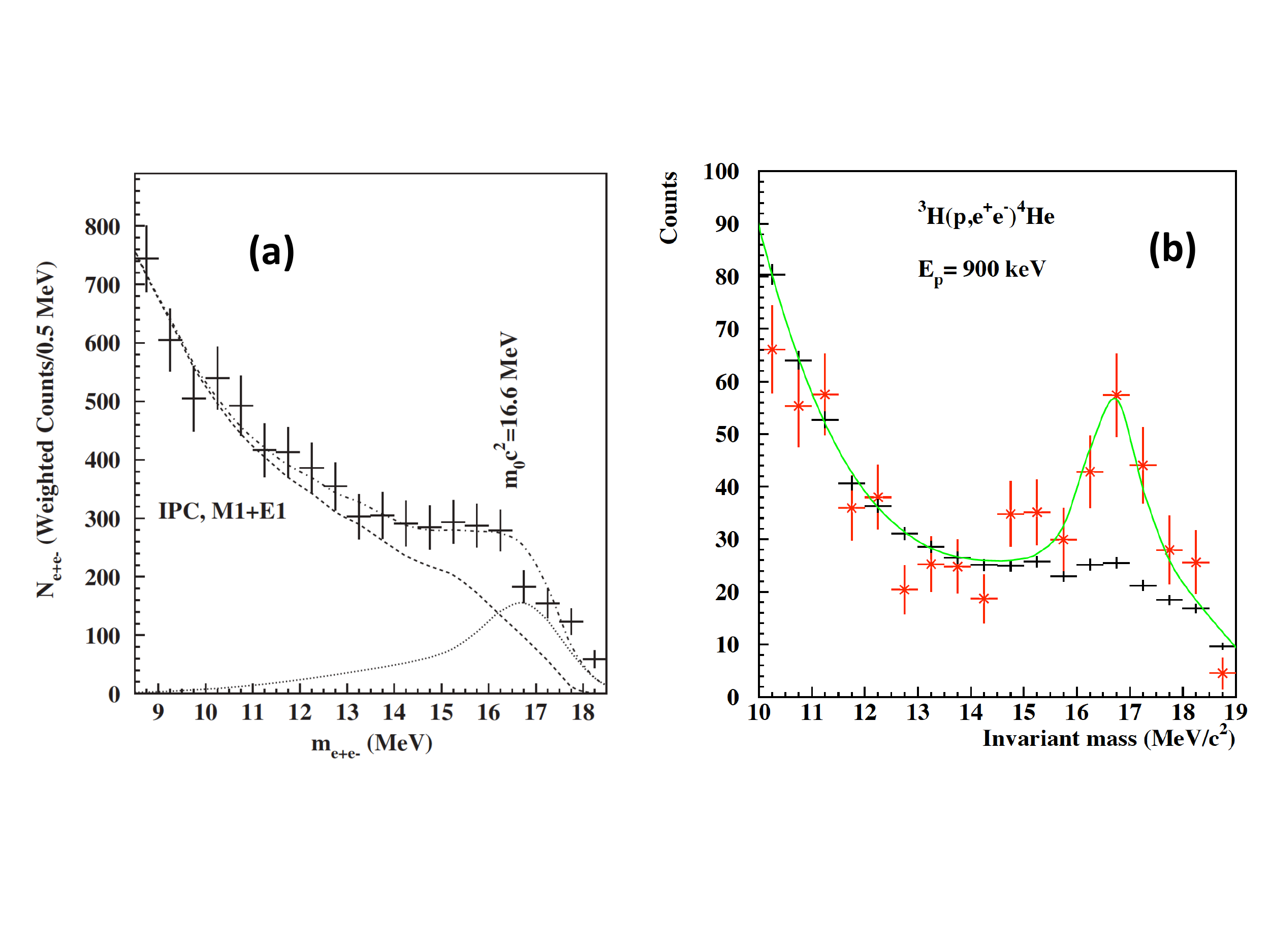} }
\caption{\label{peaks}(a): Anomaly in $^8$Be~\cite{Kra2016}. (b): Anomaly in $^4$He~\cite{Kra2019}.}
\end{figure}

It has been shown that different couplings to quark and lepton flavors could reconcile the ATOMKI anomalies; different $u$ and $d$ couplings that produce protophobic or nearly protophobic interactions would satisfy current limits~\cite{Fen2016}.  In light of that observation, the X17 anomaly has been interpreted in various more specific theoretical models as a new particle, a Z', axion, or other light pseudoscalar~\cite{Ell2016, Alv2017,Rose2019a,Rose2019b}. There are also several proposed explanations for the X17 particle within the standard model framework arising from higher order effects~\cite{Zhang2017,Ale2021,Kalman2020}.

\section{Experiment Design}

The beam energy at the ARIEL e-linac is relatively low, currently \SI{31}{MeV}, but will be upgraded to \SI{50}{MeV}. The advantage of the low beam energy is the smaller boost given to the produced $e^+e^-$ pairs from the decay of the X17 boson. The smaller boost corresponds to a large opening angle. The proposed experiment takes advantage of these larger angles.  DarkLight will run with a two spectrometer setup, optimized for the 17 MeV/$c^2$ invariant mass region. The spectrometers will be placed asymmetrically around the fixed foil target located in the beam line. The proposed experiment will measure the process  $e^-\mathit{Ta}\ \rightarrow e^-\mathit{Ta}X \rightarrow e^-\mathit{Ta}\ (e^+e^-)$ as a resonant excess of $e^+e^-$ pairs on top of the QED background at the invariant mass of the X17. 

\subsection{The Electron Accelerator}

\begin{figure}[h!]
\centering \includegraphics[width=0.7\textwidth]{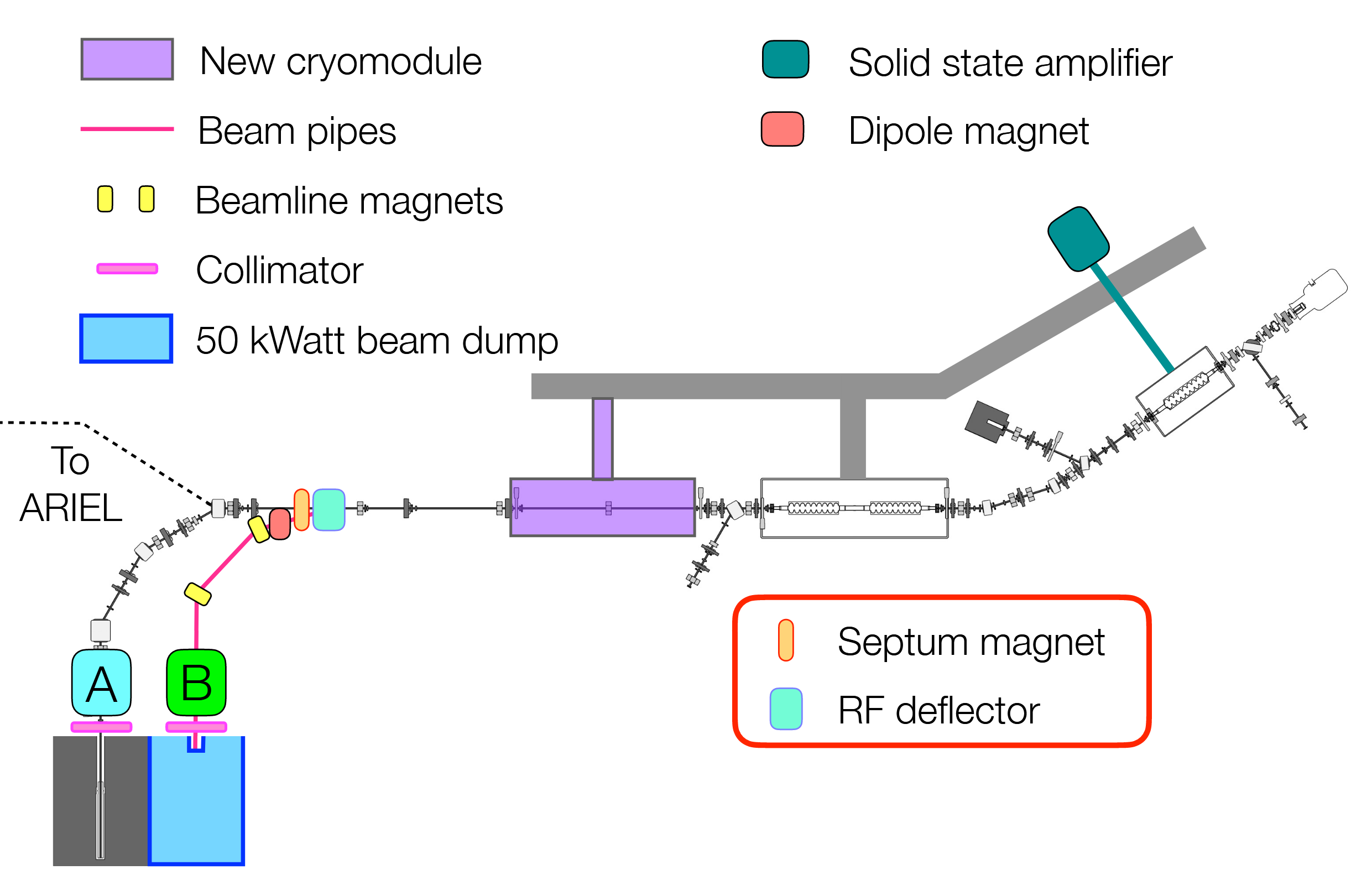}
\caption{Floorplan of the ARIEL linac. For the first measurements at \SI{31}{MeV}, the experiment will be placed in the area marked A. For \SI{50}{MeV} beam, it will be moved to position B. The septum magnet and RF deflector enables concurrent beam delivery to ARIEL and DarkLight.}
\label{triumf2}
\end{figure}

TRIUMF's existing superconducting electron linac can currently produce an electron beam of up to 31\,MeV in energy and peak intensities up to 3\,mA. As a driver of ARIEL the e-linac is designed to deliver electrons to a photo-converter target station for the production of neutron-rich rare isotope beams via photo-fission. For testing and production running at 31 MeV, the experiment will be placed in front of the existing \SI{10}{kW} beam dump (position A in Fig.~\ref{triumf2}). The linac will be operated in a continuous wave mode, with a bunch frequency of 650 MHz, and an average beam current of up to \SI{300}{\micro\ampere}. As this configuration involves minimal modifications to the beam line, data taking could commence in 2023.

The TRIUMF e-Linac is planning an upgrade to the overall beam energy deliverable to ARIEL. The approach to reach \SI{50}{MeV} involves the installation of a second superconducting cavity and cryomodule. The cavity could be installed early/mid 2024, and production data taking for DarkLight at 50 MeV can commence late 2024.

A new \SI{50}{kW} beam dump will be installed, and the experiment will move to Position B in Fig.~\ref{triumf2}. In order to facilitate simultaneous data taking with DarkLight and the rare isotope laboratory, a septum magnet and RF deflector will be added. The increased beam energy available will allow DarkLight to perform a comprehensive search of the parameter space of the X17.

\subsection{Target}

The experiment design assumes several beam energies, ranging from \SI{30}{\mega\electronvolt} to \SI{50}{\mega\electronvolt} with a current of \SI{150}{\micro\ampere}. It will impinge on a \SI{1}{\micro\metre} tantalum foil. This produces an instantaneous luminosity of $\mathcal{L}=\SI{5.2}{\per\nano\barn\per\second}$, and will cause a beam spread of approximately \SI{0.5}{\degree} 
downstream of the target.

The beam will heat up the foil with about \SI{0.4}{\watt}, which can be dissipated via radiation for typical beam spot sizes. Also under consideration is a spinning foil disc target, linked to the accelerator Fast Shutdown, to protect against accidental melting.

\subsection{Spectrometers}

The experiment will measure final state $e^+e^-$ pairs using two dipole spectrometers, with very similar magnetic characteristics, to be built by MIT/Bates. The magnetic design and pole shapes of the spectrometer have been completed. Currently the mechanical design of the supports and coils are being finalized. Both spectrometers will be designed to nearly the same specifications, presented in Table \ref{tabdesign}.

\begin{table}[!htb]
\caption{\label{tabdesign}Design parameters for the spectrometers.}
\begin{center}
\lineup
    \begin{tabular}{@{}lr@{}}             
         \hline
         \hline
         In-plane acceptance & $\pm2^\circ$\\
         Out-of-plane acceptance & $\pm$\SI{5}{\degree}\\
         Momentum acceptance & $\pm20\,\%$ \\
         Minimum central angle    & \SI{16}{\degree} \\
         Maximum central momentum & \SI{28}{\mega\electronvolt}\\
         Dipole field & \SI{0.32}{\tesla} \\
         Nominal bend radius & \SI{30}{\centi\metre}\\
         Pole gap & \SI{4}{\centi\metre}\\
         \hline
         \hline
    \end{tabular}
\end{center}
\end{table}

\begin{figure}[tb]
{\centering 
\includegraphics[width=\linewidth,clip,trim=0 1cm 0 2cm]{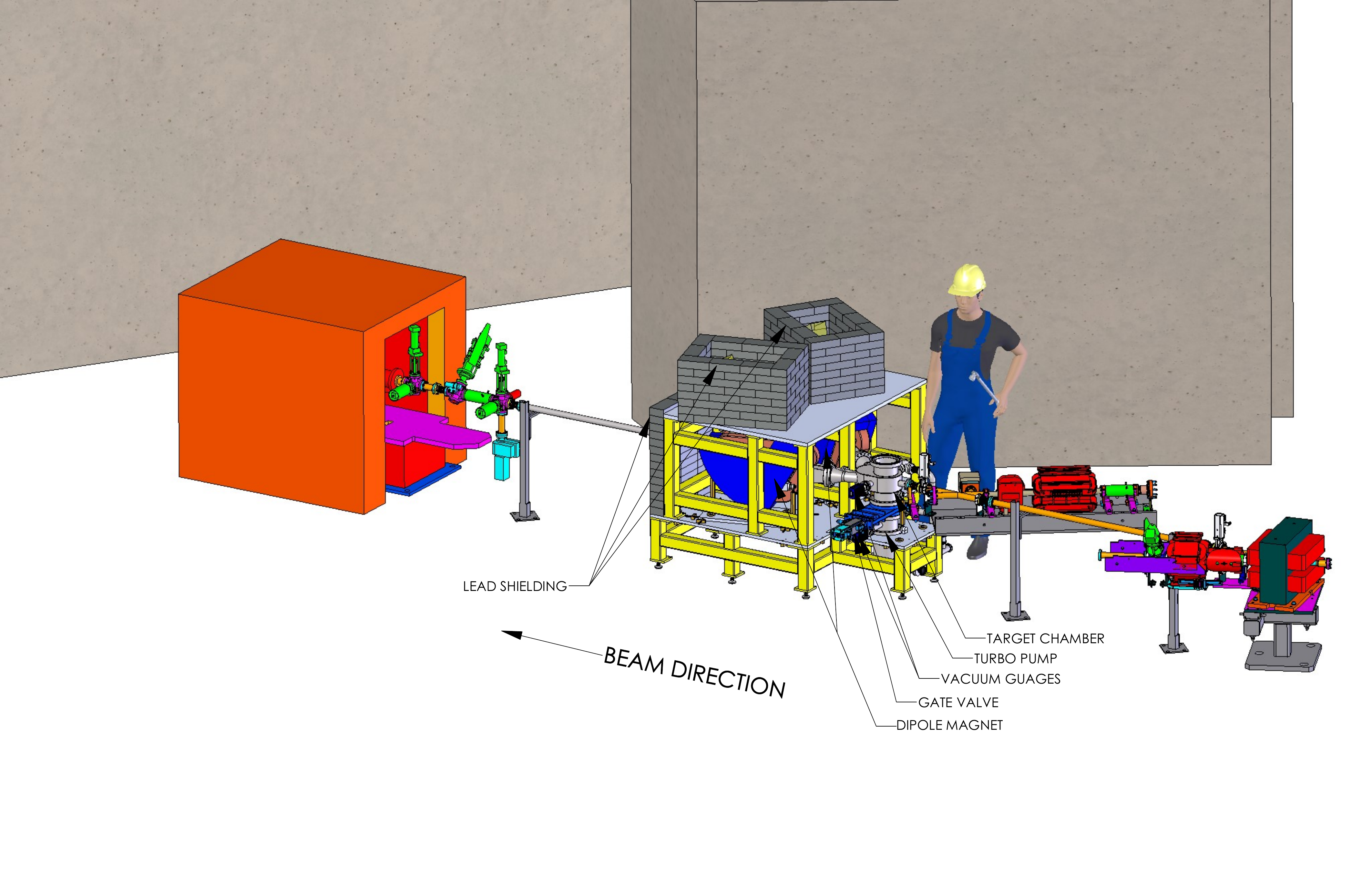}}
\caption{\label{fig:3dview} 3D CAD rendering of the conceptual design, with part of the shielding.  Additional shielding around the target is anticipated. The exit beam line will be conical (6 cm radius at 3 m distance) to allow for the increased beam width from the target interaction.}
\end{figure}

The two spectrometers share a common design but will be operated at different currents to produce the desired magnetic fields. They are conventional iron-core magnets with simple, planar coils. The magnet design and pole face rotations were optimized for a \SI{0.5}{\metre} distance from target to spectrometer entrance and for post-magnet trajectories suitable for tracking with 40~cm long GEMs. 
The final engineering of the magnet will include detailed design optimization to increase magnetic performance, minimize size, and maximize clearance to the exit beamline. The magnet in its present configuration weighs about 950\,kg.  The magnets will have full fiducialization to allow for laser tracking alignment and a six-strut mechanical support system to allow for \SI{200}{\micro\metre} alignment (similar to other MIT-Bates designs). 
A 3D CAD rendering of the experiment and spectrometers is shown in Fig.\ \ref{fig:3dview}. The focal plane of each spectrometer will be instrumented with triple-GEM detector planes followed by trigger scintillators.

\textbf{GEM detectors:} Each spectrometer will be instrumented with an identical tracking detector 
system consisting of triple-GEM elements with an active area of 25x40~cm$^2$, which have been built by Hampton University.  
These modules have two-dimensional APV front-end readout cards with 400\,$\mathrm{\mu}$m pitch between strips. The APVs are read out into Multi-Purpose Digitzer front-end cards.
They were constructed using the so-called NS2 scheme~\cite{NS2}.
A similar system of these GEMs+APVs+MPDs has recently been mass-produced for the Super-Bigbite Spectrometer (SBS) construction at Jefferson Lab. The existing GEMs can be tested and commissioned with cosmics within 9-12 months.

\textbf{Trigger Hodoscopes:} The standard GEM readout requires a trigger signal. This will be generated from the coincidence of two fast trigger detectors in the spectrometers. 
To reduce accidental coincidences in the trigger logic, it is important to resolve the beam bunch clock of \SI{650}{\mega\hertz} in the analysis. This high-resolution timing information must be provided by the trigger detector. When performing offline analysis, the timing can be corrected by the particle path length reconstructed from the tracking detector information. However, to reduce readout dead-time and data volume, it is important to be close to the ideal timing during data-taking. The main time dispersion is generated by the momentum-dependent dispersion inside the spectrometers. We therefore propose a trigger detector made from scintillator paddles, divided along the dispersive direction into 10 segments, each end read out by SiPMs.  

The scintillator paddles will be made from a standard plastic scintillator material and have a size of about 150x30x2\,mm$^3$.

\section{Projected Reach}

The shape of the background is dominated by the overall acceptance, so the irreducible background and random coincidences similar. To estimate the total background, we can scale the irreducible background up according to the predicted rates. The reach is calculated by integrating the background over the expected signal width ($\pm 1.7 \sigma$), and calculating the fifth force coupling ($\epsilon^2$) such that the signal would be bigger than a 2$\sigma$ fluctuation of the background. 

If $\epsilon^2$ is small enough the signal may not be visually detectable on top of the background. Given the excellent statistical precision of the experimental background it will still be possible to detect fluctuations corresponding to a signal in the analysis.

Since random coincidences dominate the background, the pure random background will be very accurately measured by the experiment itself by mixing electron and positron spectrometer data from different events. This mixing destroys all correlations between the spectrometers, generating a pure sample of the random coincidences. Since, in principle, every combination of uncorrelated events can be used, the available statistics for the background measurement grows quadratically with the recorded number of events. 

It is worth noting that at the kinematics and beam conditions of the experiment, the random coincidence background dominates, and scales with $\mathcal{L}^2$. The figure of merit (FOM) is given by the number of signal events divided by the square root of the background. This means in a regime where the random background dominates the FOM is independent of the instantaneous luminosity. The reach of the experiment depends only on the measurement time.

\begin{figure}[h!]
{\centering
\includegraphics[width=0.95\textwidth]{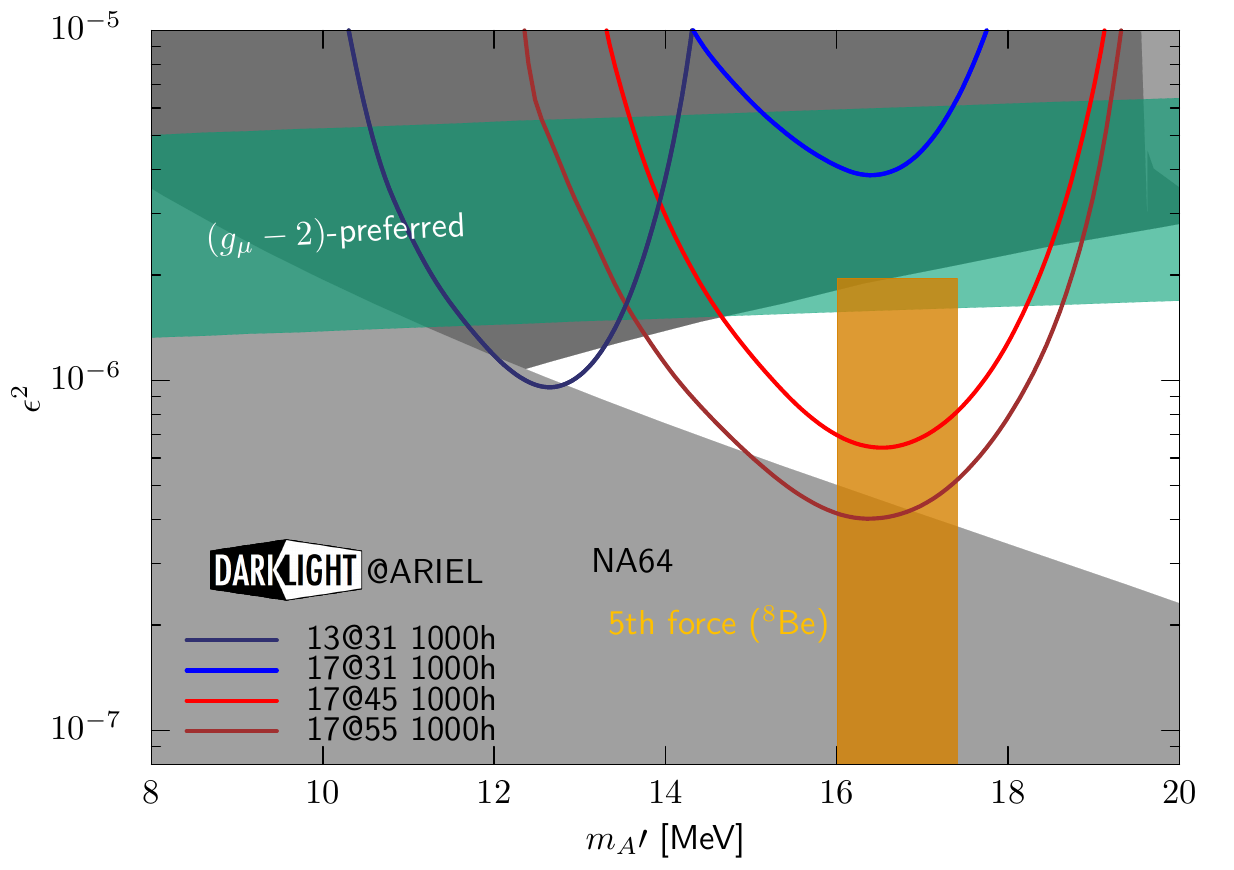} }
\caption{The projected reaches on a linear plot for three separate data taking runs: 13@31 (dark blue) $-$ a 1000~h run at 31~MeV optimized for $m_{A^\prime}$ = 13~MeV; 17@31 (light blue) $-$ a 1000~h run at 31~MeV optimized for $m_{A^\prime}$ = 17 MeV; 17@45 (light red) $-$ a 1000 h run at 45~MeV optimized for $m_{A^\prime}$ = 17~MeV; 17@55 (dark red) $-$ a 1000~h run at 55~MeV optimized for $m_{A^\prime}$ = 17~MeV. Light gray areas are excluded by other experiments sensitive to a lepton coupling. The dark gray area is excluded by electron g-2 only.}
\label{reach1}
\end{figure}

Figure \ref{reach1} shows the achievable reach for the four settings assuming 1000 hours beam-time each. The precise optimization of the experimental reach is still in progress.

\section{Summary}
As described above, the DarkLight experiment is designed to seek the protophobic new force suggested by the ATOMKI anomalies. The experiment represents a natural evolution of a fixed-target experiment to reach the low mass expected if those anomalies truly represent a new particle.  By using a high quality, low energy beam, the experiment simplifies the final states where only $e^+e^-$ pairs, with no hadronic backgrounds detected. The experiment can operate at the instantaneous luminosity saturation point, where quadratic backgrounds dominate, and the FOM depends only on measurement time. The initial 1000 hour experimental run will seek a 13\,MeV/$c^2$ resonance and explore a small region of g-2 favored parameter space beginning in 2023, limited by the beam energy currently available at ARIEL.  The apparatus will require only minor adjustment of the angles and magnet currents to accommodate the higher energies from the proposed ARIEL upgrades, which will enable searching in the 17\,MeV/$c^2$ range directly.

\section{Acknowledgements}
This work was supported by the National Science Foundation grant number PHY-2012114.  We would also like to thank the Gordon and Betty Moore Foundation for their generous support of this workshop.

\end{document}